\documentclass[11pt,a4paper]{article}
\usepackage[hyperref]{emnlp2020}
\usepackage{times}
\usepackage{latexsym}
\usepackage{multirow}
\usepackage{amsmath, amssymb}
\usepackage{graphicx}
\usepackage{subfigure}
\usepackage{float}
\usepackage[boxed,ruled,lined]{algorithm2e}
\usepackage{algorithmic}
\usepackage{array}
\usepackage{booktabs}
\usepackage{longtable}
\usepackage{tabularx}

\usepackage{microtype}

\aclfinalcopy 

\title{Context Reinforced Neural Topic Modeling over Short Texts}

\author{
	Jiachun Feng$^1$, Zusheng Zhang$^1$, Cheng Ding$^1$,\\ {\bf Yanghui Rao$^1${\thanks{ ~\ The corresponding author.}}, Haoran Xie$^2$} \\
	$^1$School of Data and Computer Science, Sun Yat-sen University, Guangzhou, China \\
	$^2$Department of Computing and Decision Sciences, Lingnan University, Hong Kong \\
	{\tt \{fengjch5, zhangzsh3, dingch6\}@mail2.sysu.edu.cn,} \\
	{\tt raoyangh@mail.sysu.edu.cn, hrxie2@gmail.com}
}

\begin{document}
\maketitle
\begin{abstract}
As one of the prevalent topic mining tools, neural topic modeling has attracted a lot of interests for the advantages of high efficiency in training and strong generalisation abilities. However, due to the lack of context in each short text, the existing neural topic models may suffer from feature sparsity on such documents. To alleviate this issue, we propose a Context Reinforced Neural Topic Model (CRNTM), whose characteristics can be summarized as follows. Firstly, by assuming that each short text covers only a few salient topics, CRNTM infers the topic for each word in a narrow range. Secondly, our model exploits pre-trained word embeddings by treating topics as multivariate Gaussian distributions or Gaussian mixture distributions in the embedding space. Extensive experiments on two benchmark datasets validate the effectiveness of the proposed model on both topic discovery and text classification.
\end{abstract}

\section{Introduction}
Mining topics from texts is significant for various applications of natural language processing, e.g., text classification, sentiment analysis, and recommender systems. As one of the most popular approaches for discovering latent topics, topic modeling \cite{LDA, DMM} is capable of producing interpretable results. Generally, the dominant methods for parameter estimation in topic models are variational inference \cite{LDA} and Gibbs sampling \cite{Gibbs}, both of which, however, require complex re-derivation when there is any minor changes to the model structure. Moreover, with the growth of data scale, the generative process is getting tricky and expensive, which leads to mathematically arduous derivation and high computational cost in training. These limitations make it difficult to extend the models to new variations flexibly.

With the development of deep learning, variational auto-encoder (VAE) \cite{VAE} has provided another promising solution for topic modeling. Benefiting from the flexibility of neural networks, the VAE framework is competent to learn complicated non-linear distributions and is convenient to be applied to various tasks. Furthermore, by using the back-propagation for optimization, VAE is highly efficient in training when compared with the models based on variational inference or Gibbs sampling. Considering the above advantages, several models built on VAE have been proposed, such as neural variational document model (NVDM) \cite{NVDM}, neural variation latent Dirichlet allocation (NVLDA) \cite{NVLDA}, Gaussian softmax model (GSM) \cite{GSM}, Dirichlet variational auto-encoder (DVAE) \cite{burkhardt2019decoupling}, and neural variational correlated topic modeling (NVCTM) \cite{NVCTM}. Although the VAE-based models reduce the computational cost impressively, they still suffer from the feature sparsity problem in short texts. In this case, the number of word occurrences in each text is relatively small, while the vocabulary corresponding to the corpus is large and the range of topics is broad.

To alleviate the above issue, many Bayesian approaches specific to short texts have been proposed \cite{DBLP:conf/www/YanGLC13, DBLP:conf/www/LinTMC14, GPUDMM}. Nonetheless, the above models all resort to Gibbs sampling or variational inference and hence incur the problems as mentioned before. In recent years, models built on VAE are also introduced for short texts, such as Graph-based inference network for the biterm topic model (GraphBTM) \cite{GBTM} and neural sparsemax topic model (NSMTM) \cite{DBLP:conf/wsdm/LinHG19}. However, learning context information is still challenging in these models due to significant word non-overlap in short texts. Relatedness information between word pairs may not be fully captured owing to the lack of word-overlap between such short messages.

In this paper, we propose a VAE-based topic model for short texts, where the context information for each text is effectively enhanced. Firstly, as can be observed, a short text generally covers only a subset of topics due to the limited text length. Therefore, we propose to filter irrelevant topics by setting a \emph{topic controller} for each topic, encouraging each short text to focus on some salient topics. Through this way, the topic inference range is narrowed down and thus the topic sparsity can be achieved indirectly. Secondly, we incorporate pre-trained word embeddings into our model to explicitly enrich the context information. Specifically, we model each topic by a multivariate Gaussian distribution or a Gaussian mixture distribution in the embedding space, through which the relatedness of synonymous word pairs can be effectively inferred regardless of word non-overlap in short texts. In this way, our model can discover more interpretable topics than other topic models. We name the proposed model as Context Reinforced Neural Topic Model (CRNTM) and conclude the main contributions of our work as follows:
\begin{itemize}
\item We assume that each short text only focuses on a few salient topics. By setting a \emph{topic controller} for each topic to filter irrelevant topics, CRNTM narrows down the topic inference space and achieves topic sparsity indirectly.
\item Pre-trained word embeddings are incorporated to explicitly enrich the limited context information for each short message. By treating topic distributions over words as multivariate Gaussian distributions or Gaussian mixture distributions in the embedding space, CRNTM can produce more interpretable topics.
\end{itemize}

The rest of this paper is organized as follows. We discuss relevant research work in Section \ref{sec:relatedwork}, and detail our proposed model in Section \ref{sec:crntm}. Experimental settings and results are presented in Section \ref{sec:experiment}. Finally, we draw the conclusion in Section \ref{sec:conclusion}.

\section{Related Work} \label{sec:relatedwork}
\subsection{Neural Topic Modeling}
With the development of deep learning, models built on neural networks have been proposed to discover latent topics, and most of them are based on VAE. In this vein, NVDM \cite{NVDM} is a neural variational framework for generative modeling on texts. It consists of an inference network and a multinomial softmax generative module. The inference network is used to estimate continuous hidden variables, which can represent the semantic content of documents, while the generative module aims to reconstruct the documents from the latent topic distributions. GSM \cite{GSM} constructs the topic distributiona explicitly with a softmax function applied to the projection of the Gaussian random vector. ProdLDA \cite{NVLDA} replaces the mixture model in latent Dirichlet allocation (LDA) with a product of experts for better topic modeling. NVLDA \cite{NVLDA} approximates the Dirichlet distribution by using Laplace approximation. DVAE \cite{burkhardt2019decoupling} decouples the properties of sparsity and smoothness by rewriting the Dirichlet parameter vector into a product of a sparse binary vector and a smoothness vector. NVCTM \cite{NVCTM} enhances the capability of capturing the correlations among topics by reshaping topic distributions.

\subsection{Short Text Topic Discovery}
Topic models \cite{LDA} provide a valuable solution for implicit semantic mining and understanding over documents. However, the feature sparsity problem arises for topic models when applied to short texts \cite{TWLDA}, because such corpora are lack of word co-occurrences at the document level.

To overcome this limitation, the external documents were first introduced to enrich the contextual information in short texts \cite{7,12,13}. Unfortunately, it requires the external documents to be semantically close to the original corpus. Some approaches tackle the task by aggregating short texts into lengthy pseudo-documents and then applying a well established topic model. For this category of methods, short texts can be aggregated by utilizing the side information, e.g., user characteristics tags \cite{UGTE}, user ID \cite{TWLDA}, and timestamp \cite{TURLDA}. Another alternative methods directly modify the prior of Bayesian models to enrich word co-occurrences, so as to remedy the feature sparsity problem. For instance, the Biterm Topic Model (BTM) \cite{BTM}, which models the global word co-occurrences at the corpus level, could lengthen short texts by converting documents into biterm sets.

While the above methods are developed based on Bayesian models, some neural network based approaches have been introduced for short texts. \citet{GBTM} proposed a graph-based inference network named GraphBTM for accelerating the above BTM. This model sampled a fixed number of texts as a training instance to overcome the feature sparsity issue. \citet{DBLP:conf/wsdm/LinHG19} proposed a neural model which is called NSMTM by providing sparse posterior distributions over topics based on the Gaussian sparsemax construction. \citet{DBLP:conf/aaai/GuptaCBS19} designed a neural autoregressive topic model named iDocNADE in a language modeling fashion. They also incorporated word embeddings as fixed prior in the model to introduce complementary information. However, the above approach does not model topic distributions explicitly.

\section{Model Description} \label{sec:crntm}
\begin{figure}[h]
  \centering
  \includegraphics[scale=0.22]{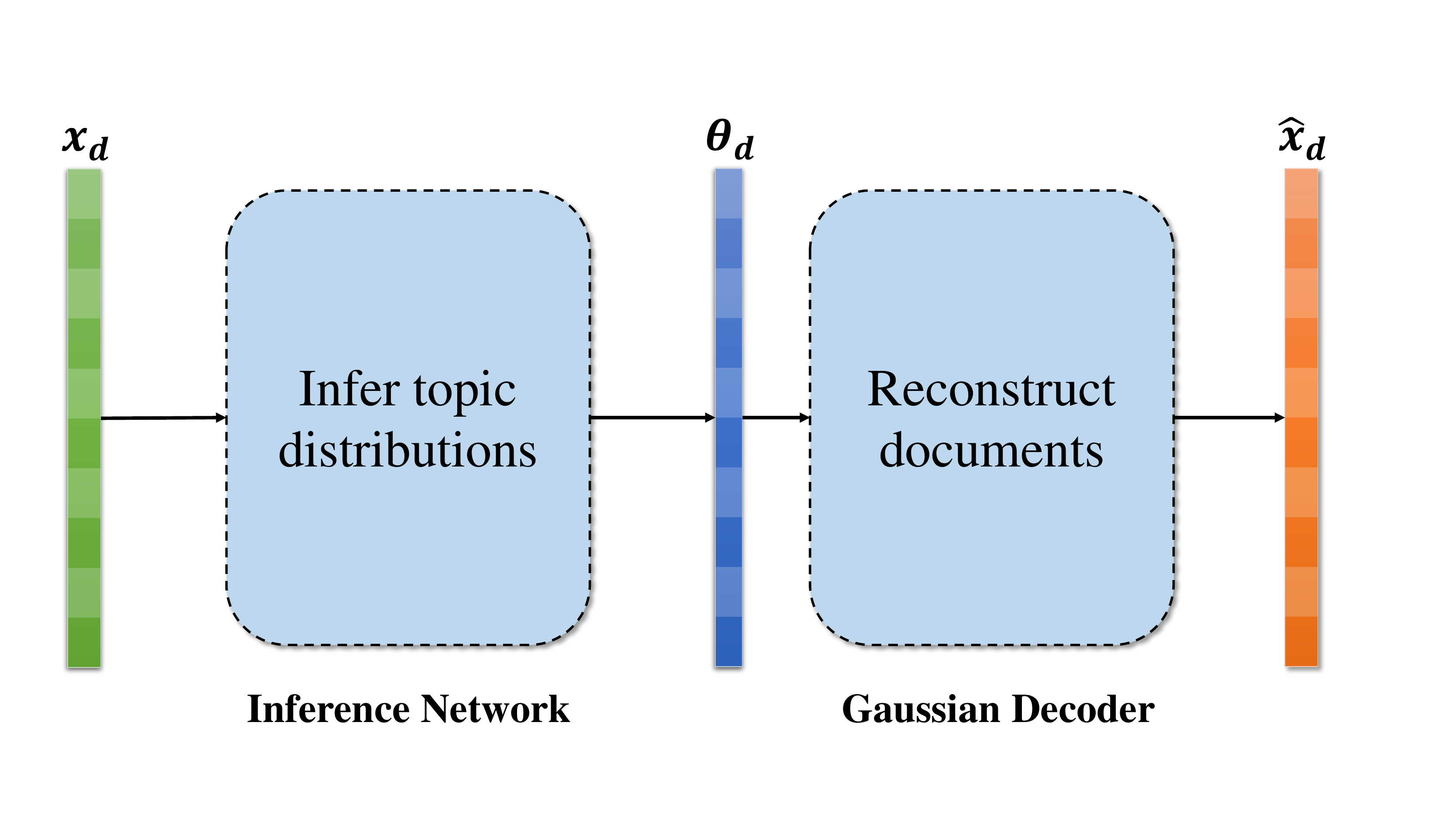}
  \caption{Structure of our CRNTM.}\label{fig:CRNTM}
\end{figure}

In this section, we describe our context reinforced neural topic model (CRNTM) in details. The overall architecture is illustrated in Figure \ref{fig:CRNTM}, which consists of two major modules: an inference network for learning latent topics, and a Gaussian decoder for reconstructing documents.

\subsection{Problem Definition}\label{sec: crntm-questionDescribe}
Given a corpus with $D$ short texts, we denote the corresponding vocabulary as $ W = \{w_1, w_2, ..., w_{V} \} $, with $V$ being the vocabulary size. Following \cite{NVDM}, each document  is processed into a bag-of-words (BOW) representation, i.e., $ x_ {d} = [x_ {d, 1}, x_ {d, 2}, ..., x_ {d, V}] $,  where $ x_ {d, i} $ denotes the number of times for word $ w_i $ appearing in  document $d$.

In the inference network, we use $\theta_d \in \mathbb{R}^K$ to denote the topic distribution of document $d$ and use $z_k \in \{z_1, z_2, ..., z_{K} \}$ to denote the topic assignment for an observed word, where $K$ denotes the number of topics inherent in the given corpus. Specifically, $\theta_d$ is drawn from the Gaussian distribution $\mathcal{N} (\mu_d, \Sigma_d)$, where both $\mu_d$ and $\Sigma_d$ are prior parameters. Furthermore, we set a \emph{topic controller} $\lambda_{d,k} \in [0,1]$ for each topic $z_k$ in document $d$: the topic will be kept when $\lambda_{d,k}=1$, or it will be filtered out when $\lambda_{d,k}=0$. For document $d$, the \emph{topic controller} $ \lambda_ {d} = \{\lambda_ {d, k} \} _ {k = 1} ^ {K} $ is drawn from a Beta distribution, i.e., $ \lambda_ {d} \sim \mathrm{Beta} (\alpha_d, \beta_d) $, where $\alpha_d$ and $\beta_d$ are the prior parameters of $\lambda_d$.

For the Gaussian decoder, we denote the word embedding matrix corresponding to the vocabulary as $WE \in \mathbb{R}^{V \times r}$, where $r$ indicates the dimension of word embeddings. Moreover, the embedding of word $w_i$ is represented as $WE_i$. We use a matrix $TW \in \mathbb{R}^{K \times V}$ to denote the probabilities of words conditioned to topics, in which, $TW_{(k,i)}$ represents the conditional probability of word $w_i$ over topic $z_k$. In this study, $TW_{(k,i)}$ is drawn from a multivariate Gaussian distribution $\mathcal{N} (\mu_k, \Sigma_k)$ or a Gaussian mixture distribution, where $\mu_k \in \mathbb{R}^r$ and $\Sigma_k \in \mathbb{R}^{r \times r}$ are learnable parameters.

\subsection {Inference Network}
\begin{figure}[h]
  \centering
  \includegraphics[scale=0.22]{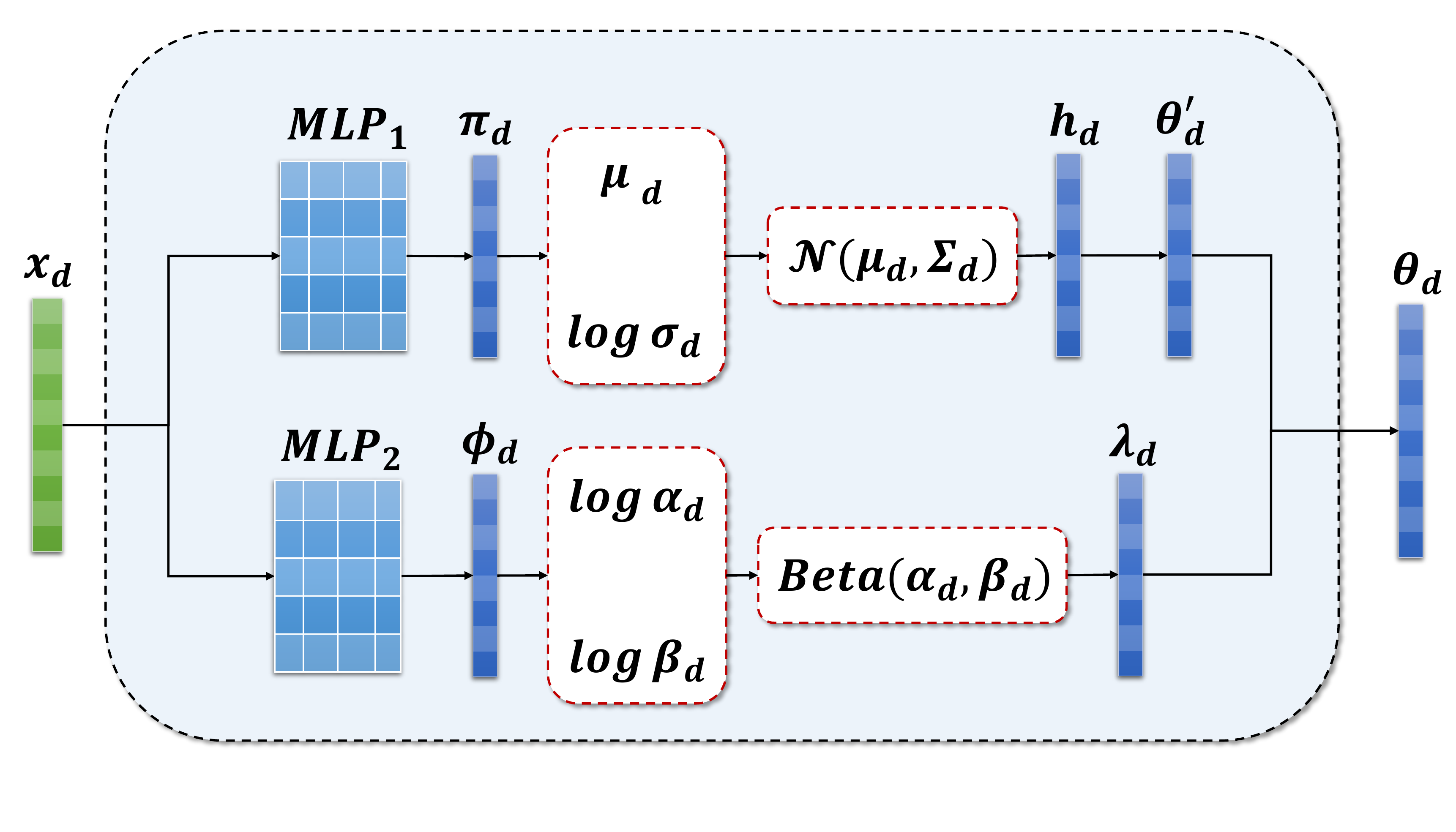}
  \caption{Inference network.}\label{fig:infnet}
\end{figure}

The first component of CRNTM is the inference network, which is applied to infer the topic distributions for the input documents.  The structure of our inference network  is illustrated in Figure \ref {fig:infnet}. Following the framework of VAE, CRNTM infers the parameters $\mu_d$ and $\Sigma_d$ via deep neural networks that are elaborately designed for the observed data. Being fed with the input document $ x_d $, the inference network first outputs an encoded vector $ \pi_d $. Then, $ \pi_d $  is linearly transformed to obtain $  \mu_d$ and  $\Sigma_d $,  which are used to parameterize the Gaussian prior $ \mathcal {N} (\mu_d, \Sigma_d) $. The above process is described by $\pi_d = \mathrm{MLP_1} (x_d)$, $\mu_d = l_1(\pi_d)$, $\log \sigma_d = l_2(\pi_d)$, and $\Sigma_d = \mathrm{diag} (\sigma_d^2)$,
where $ \mathrm{MLP_1} $ is a multilayer perceptron, $ l_1 (\cdot) $ and $ l_2 (\cdot) $ are linear transformations. Note that the diagonal elements $ \sigma_d^2 $ of covariance matrix $ \Sigma_d $ are non-negative. The output of $ l_2 (\cdot) $ is regarded as the logarithmic form $ \log \sigma_d $, which is a real number.

A Gaussian random vector $h_d$ is passed through a softmax function to parameterize the multinomial document-topic distribution $\theta_d'$. The process is defined as: $\epsilon_d \sim \mathcal{N} (0, I^2 )$, $h_d = \mu_d + \epsilon_d * \sigma_d$, and $\theta_d' = \mathrm{softmax}(W_{\theta}  \cdot  h_d + b_{\theta})$,
where $ h_d $ is drawn from the Gaussian prior $ \mathcal {N} (\mu_d, \Sigma_d) $ with re-parameterization, allowing the parameters to be optimized by back-propagation.

Due to the limited text length, a short document only contains a few of words, resulting in the feature sparsity problem during the inference process. However, it can be observed that a short text generally focuses on a subset of topics. This inspires us to alleviate the above issue by narrowing down the scope for topic inference. Instead of letting the topic mixtures navigate freely in the simplex, CRNTM allows short texts to cover a narrow range of topics. This is achieved by setting a \emph{topic controller} $ \lambda_ {d, k} \in [0, 1] $ for each topic. Topic $z_k$ is focused when $\lambda_{d, k}=1$, and it will be filtered out when $\lambda_{d, k}=0$. The topic controllers are drawn from the Beta distribution, i.e., $ \lambda_{d} \sim \mathrm{Beta} (\alpha_{d}, \beta_{d}) $, as described in Section \ref{sec: crntm-questionDescribe}, which can guarantee that each component $\lambda_{d,k} \in \lambda_d$ is in the range of $[0,1]$. The parameters $\alpha_{d}$ and $\beta_{d}$ are inferred as follows:
\begin{eqnarray}
\phi_d & = &  \mathrm{MLP_2} (x_d), \label{eq:phi}\\
\log \alpha_d  & = &  l_3(\phi_d), \label{eq:palpha}\\
\log \beta_d & = & l_4(\phi_d), \label{eq:pbeta}
\end{eqnarray}
where $ \mathrm{MLP_2} $ is a multilayer perceptron, $ l_3 (\cdot) $ and $ l_4 (\cdot) $ are linear transformations. The output of  $ l_3 (\cdot) $ and  $ l_4 (\cdot) $ are treated as the logarithmic form  $\log \alpha_{d}$ and $\log \beta_{d}$, since both $\alpha_d$ and $\beta_d$ are non-negative.

The Beta sampling can not be  differentiated directly,  making it intractable to update model parameters through back-propagation. Therefore, we use the  re-parameterization technique to obtain $ \lambda_{d} $ by following \cite{RepBeta}. The sampling operation of $ \mathrm{Beta} (\alpha_d, \beta_d) $  can be decoupled into $ \mathrm{Gamma} (\alpha_d, 1) $  and $ \mathrm{Gamma} (\beta_d , 1)$, which is formulated by $\lambda_{d} = \frac{\lambda_{d,1} }{\lambda_{d,1}  +\lambda_{d,2}}$,
where $\lambda_{d,1} \sim  \mathrm{Gamma} (\alpha_d, 1) $ and $ \lambda_{d,2} \sim \mathrm{Gamma} (\beta_d , 1)$. For the Gamma distribution $ \mathrm{Gamma} (\alpha, 1) $ with $\alpha >  1$, the re-parameterization can be accomplished by the reject sampling method:
\begin{equation}
\lambda_{d,1} =   (\alpha_d-  \frac{1}{3} ) ( 1 + \frac{\epsilon_d}{\sqrt{9 \alpha_d -3}})^3, \label{eq:alpha>1}
\end{equation}
where $\epsilon_d \sim  \mathcal{N}(0, I^2) $. On the other hand, the shape augmentation method is applied to convert  $\alpha \leq  1$  to  $\alpha >  1$ to increase the accept rate of each rejection sampler, which is formulated by $\lambda_{d,1} = \rho^{\frac{1}{\alpha_d}} \tilde{\lambda}_{d,1}$,
where $\rho $ is drawn from a uniform distribution, i.e., $\rho \sim U[0,1]$, and $\tilde{\lambda}_{d,1} \sim Gamma(\alpha+1, 1)$ can be obtained according to Equation (\ref{eq:alpha>1}).

During the inference process, CRNTM determines whether topic $ z_k $ is kept according to $ \lambda_ {d, k} $. By filtering out some certain topics, the short texts are allowed to  focus on a few specific topics, and thus the feature sparsity problem can be alleviated. Finally, the topic distribution of  document $ x_d $ is obtained by $\theta_{d}= \theta_{d}' * \lambda_d$.

\subsection{Gaussian Decoder}
\begin{figure*}
\centering
\subfigure[Gaussian distribution]{
\begin{minipage}[t]{0.48\linewidth}
\centering
\label{fig:GD}
\includegraphics[scale=0.22]{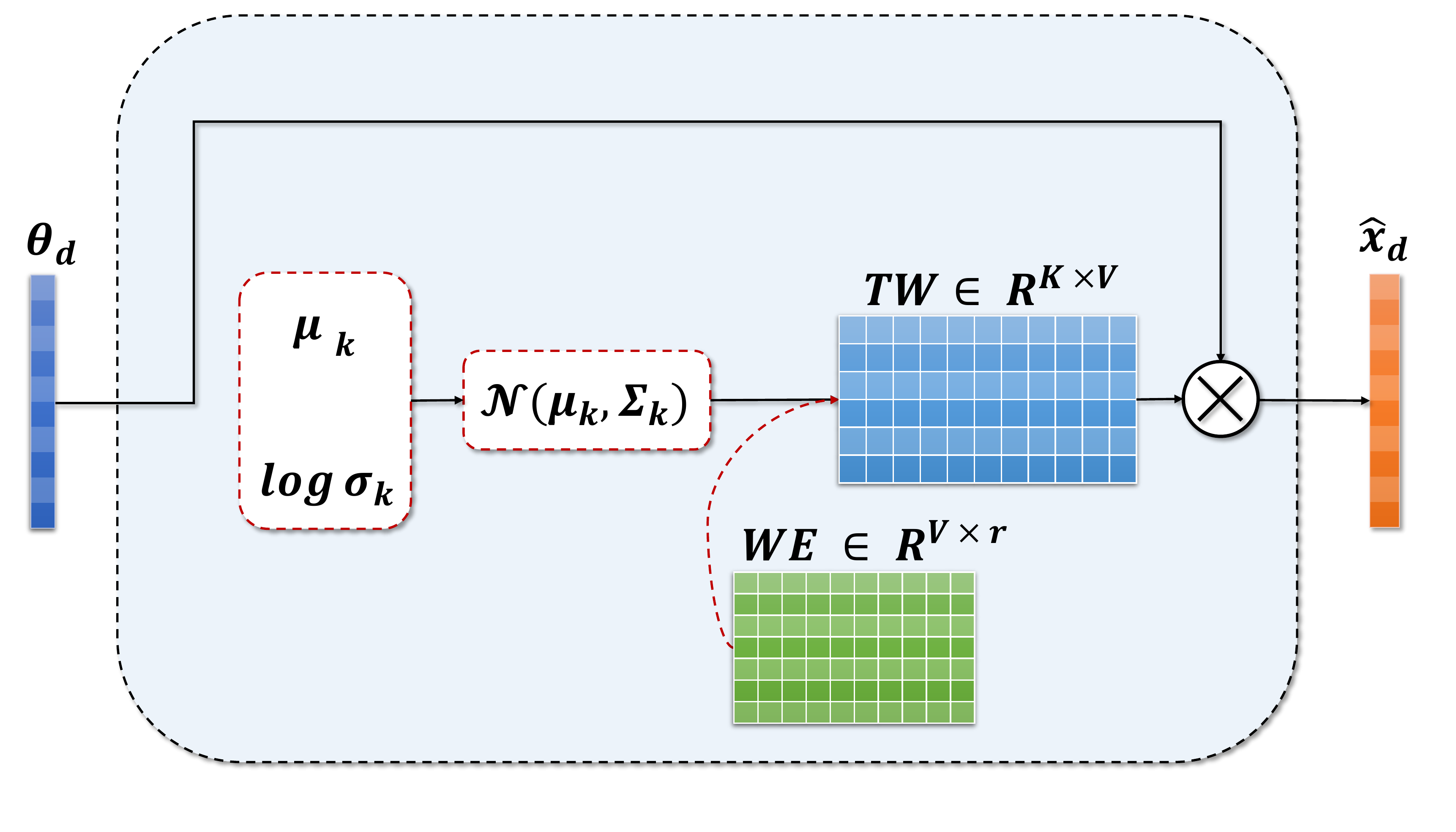}
\end{minipage}
}
\subfigure[Gaussian mixture distribution]{
\begin{minipage}[t]{0.48\linewidth}
\centering
\label{fig:GMD}
\includegraphics[scale=0.22]{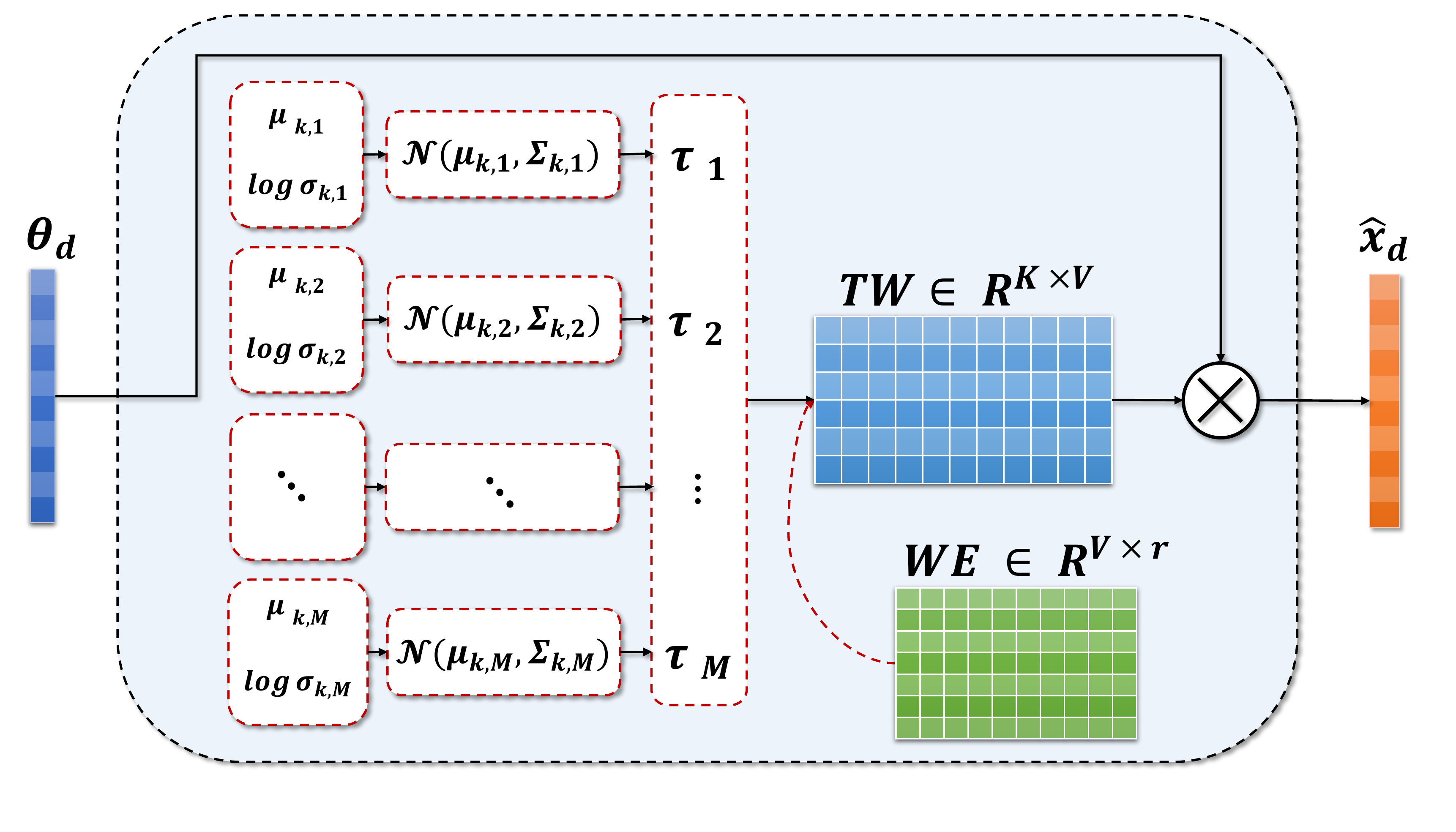}
\end{minipage}
}
\caption{Gaussian decoder.}
\end{figure*}

Context information is important for topic mining \cite{DBLP:conf/aaai/GuptaCBS19}. Words that appear together frequently are more likely to belong to the same topic, which implies that closer words in the embedding space are more likely to reflect the same topic. In Bayesian models, word embeddings that are trained on a large corpus have shown to effectively bring auxiliary context information for short texts \cite{GPUDMM}. Considering this advantage, we propose to introduce word embeddings into the decoder named Gaussian decoder. To our best knowledge, this is the first work of incorporating pre-trained word embeddings into the decoder of VAE to enhance the ability of capturing context information. The basic structure of our Gaussian decoder is shown in Figure \ref {fig:GD}.

The Gaussian decoder is applied to decode the topic distribution $\theta_d$, based of which a new document can be reconstructed. Concretely, the decoder employs the multivariate Gaussian distribution $ \mathcal {N} (\mu_ {k}, \Sigma_{k}) $ to model the $k$-th topic in the embedding space.  Since the elements of the diagonal matrix $ \Sigma_ {k} $ are non-negative, a transformation similar to the one applied to $ \Sigma_d $ is used here, as follows: $\Sigma_k = \mathrm{diag} ( \sigma_k ^2)$.

By incorporating pre-trained word embeddings, the probability of word $w_i$ conditioned on topic $z_k $, i.e., $TW_{(k,i)}$, can be formulated by $TW_{(k,i)} = \frac{\exp \left (  g(WE_i)  \right )}{(2\pi)^{r/2} \mid \Sigma_k \mid^{1/2}}$,
where $r$ is the word embedding dimension, and $g(WE_i) =  -\frac{1}{2} (WE_{i} - \mu_k)^T \Sigma_k^{-1} (WE_{i} - \mu_k)$.
It is worth noting that the parameters $\mu_k$ and $\Sigma_k$ can be regarded as the topic centroid and topic concentration in the embedding space. According to the properties of Gaussian distribution, words that are closer to the topic centroid have higher probabilities. Meanwhile, words in each topic are related to the context information implied by word embeddings. Therefore, CRNTM can enrich context information via pre-trained word embeddings to address the feature sparsity problem.
Finally, we estimate the conditional probability $p(w_{d,i}|\theta_d, \lambda_d)$ by $p(w_{d,i} | \theta_d, \lambda_d) = \sum_k \theta_{d, k} \cdot TW_{(k,i)}$.

The above method can be easily adjusted by assuming that $TW_{(k,i)}$ obeys a Gaussian mixture distribution, as shown in Figure \ref {fig:GMD}. In this case, $TW_{(k,i)}  =  \sum_{m=1}^{M} \tau_m  \frac{\exp \left (  g_m(WE_i)  \right )}{(2\pi)^{r/2} \mid \Sigma_{k,m} \mid^{1/2}}$, where $M$ is number of Gaussian components,  $\tau_m$ is the coefficient of Gaussian distributions, and $g_m(WE_i) =  -\frac{1}{2} (WE_{i} - \mu_{k,m})^T \Sigma_{k,m}^{-1} (WE_{i} - \mu_{k,m})$.

\subsection {Optimization Objective}
The optimization objective of CRNTM is $ \mathcal {L} = \log p (D) $, where $ \log p (D) $ is the likelihood of observed samples. According to the assumption, there is $ \log p (D) = \sum_d \log p (d) $. Since the true distributions of documents are unknown, variational inference is used here to convert the optimization to its evidence lower bound (ELBO), that is, $ \log p ( d) \geq \mathcal {L} (d) $. According to the variational inference method, $ \mathcal {L} (d) $ is derived as follows:
\begin{eqnarray}
\mathcal{L}(d) & = &  \iint q(\theta_d, \lambda_d | x_d)  [- \log q(\theta_d, \lambda_d | x_d)  \nonumber \\
& & +  \log p(x_d, \theta_d, \lambda_d)  ] \mathrm{d} \theta_d \mathrm{d} \lambda_d \nonumber \\
& = & E_{q(\theta_d | x_d)q(\lambda_d | x_d)}[\log p(x_d | \theta_d, \lambda_d)] \nonumber \\
& &  - D_{KL}[q(\theta_d | x_d) \parallel  p(\theta_d)] \nonumber \\
& &  - D_{KL}[q( \lambda_d | x_d) \parallel  p(\lambda_d )],
\end{eqnarray}
where $ E_ {q (\theta_d | x_d) q (\lambda_d | x_d)} [\log p (x_d | \theta_d, \lambda_d)] $ is often regarded as the reconstruction loss. $ p(x_d | \theta_d, \lambda_d) = \prod_{i=1} ^ {n_d} p (w_{ d, i} | \theta_d, \lambda_d) $. $ p (\theta_d) $ is the prior distribution of $ \theta_d $, $ q (\theta_d | x_d) $ is the variational approximation of $ p (\theta_d) $, $ p (\lambda_d) $ is the prior distribution of $ \lambda_d $, $ q (\lambda_d | x_d) $ is the variational approximation of $ p (\lambda_d) $, and $ E_ {q (\theta_d | x_d)} (\cdot) $ is approximated by sampling of $ \theta_d \sim q (\theta_d | x_d) $. For $\theta_d$, we assume that the true prior $p(\theta_d)$ is a normal Gaussian distribution $ \mathcal {N} (0, I) $ by following \cite{VAE, NVDM, NVCTM}. Therefore, the KL divergence term $D_{KL}[q(\theta_d | x_d) \parallel  p(\theta_d)] $ can be derived by $D_{KL}[q(\theta_d | x_d) \parallel  p(\theta_d)] = \frac{1}{2}( -n + \mu_d^2 - \log |\Sigma_d| + |\Sigma_d|)$.

Similarly, we take $Beta(\alpha', \beta')$ as the true prior of $p (\lambda_d) $, and the KL divergence term $ D_{KL}[q( \lambda_d | x_d) \parallel  p(\lambda_d )] $ can be computed by $D_{KL}[q( \lambda_d | x_d) \parallel  p(\lambda_d )] = \ln \frac{\Delta(\alpha', \beta')}{\Delta(\alpha_d, \beta_d)} - (\alpha' - \alpha_d) \psi(\alpha_d) -  (\beta'-\beta_d)\psi(\beta_d) + (\alpha' - \alpha_d + \beta' - \beta_d) \psi(\alpha_d + \beta_d)$,
where $\Delta(\alpha, \beta) = \frac{\Gamma(\alpha)\Gamma(\beta)}{\Gamma(\alpha + \beta)}$, $\Gamma(\cdot)$ is the Gamma function, and $\psi(\cdot)$ is the Digamma function. In our model, the \emph{topic controller} acts as a switch to filter out irrelevant topics and keep related topics with higher probabilities. Since $\alpha$ and $\beta$ determine the shape of Beta distribution, we set both $\alpha'$ and $\beta'$ to 0.5, so that the probabilities are sharp in values of 0 and 1.

\section{Experiments} \label{sec:experiment}
In this section, we first introduce the experimental setting, and then evaluate the effectiveness of our model by a series of experiments.

\subsection{Datasets}
To compare the model performance on both topic mining and text classification, we employ 20NewsGroups\footnote{\url{http://www.qwone.com/~jason/20Newsgroups/20news-18828.tar.gz}} and Snippets\footnote{\url{http://jwebpro.sourceforge.net/data-web-snippets.tar.gz}} with document labels as our datasets. 20NewsGroups is a collection of short news messages, which contains $11,314$ training and $7,531$ testing samples. These short texts are grouped into 20 different categories. Snippets is collected from the results of web search transaction over 8 domain labels. The officially divided 10,060 and 2,280 search transaction documents are used for training and testing, respectively. For data preprocessing, we remove stopwords and take the most frequent $2,000$ words and $5,000$ words as vocabularies. The statistics of the processed corpora are shown in Table \ref{tab:datasets}, where $AvgD$ and $L$ denote the averaged number of words for each document and the number of categories, respectively.

\begin{table}
  \caption{The statistics of datasets. }
  \label{tab:datasets}
\scriptsize
\begin{center}
  \begin{tabular}{cccccc}
    \toprule
    Dataset & Train   & Test  & $V$ & $AvgD$ & $L$\\
    \midrule
    20NewsGroups & $11,314$ & $7,531$ & $2,000$ & 12.3 & 20\\
    Snippets & $10,060$ & $2,280$ & $5,000$ & 14.3 & 8 \\
    \bottomrule
  \end{tabular}
\end{center}
\end{table}

\subsection{Baseline Methods}
We use the following mainstream VAE based methods as baselines for evaluation: NVDM \cite{NVDM}, NVLDA \& ProdLDA \cite{NVLDA}, GSM \cite{GSM}, TMN \cite{DBLP:conf/emnlp/ZengLSGLK18}, NVCTM \cite{NVCTM}, and DVAE \cite{burkhardt2019decoupling}. Among these methods, NVDM is one of the first neural document models, NVLDA, ProdLDA, and GSM are classical neural topic models.
TMN consists of a neural topic model and a topic memory mechanism, which are trained in an end-to-end learning manner. NVCTM exploits the Centralized Transformation Flow (CTF) to capture the topic correlations by reshaping topic distributions. DVAE achieves a competitive topic coherence and a high log-likelihood by decoupling the properties of sparsity and smoothness in VAE-based topic models for short texts.

Note that iDocNADE \cite{DBLP:conf/aaai/GuptaCBS19} and NSMTM \cite{DBLP:conf/wsdm/LinHG19} are not adopted for comparison, because the former does not model topic distributions explicitly while the training process of the latter is too sensitive to continue based on our implementation. Besides, since the ELBO is typically used and necessary to evaluate the performance of VAE based methods \cite{NVDM, GSM}, we do not use Bayesian models such as \cite{DBLP:conf/www/YanGLC13, DBLP:conf/www/LinTMC14, GPUDMM} as baselines for fair comparison. Finally, GraphBTM \cite{GBTM} which only models a mini-corpus is unsuitable to be evaluated in this study.

\subsection{Experimental Settings}
In our experiments, the publicly available codes of NVDM\footnote{\url{https://github.com/ysmiao/nvdm}}, NVLDA \& ProdLDA\footnote{\url{https://github.com/akashgit/autoencoding_vi_for_topic_models}}, TMN\footnote{\url{https://github.com/zengjichuan/TMN}}, and DVAE\footnote{\url{https://github.com/sophieburkhardt/dirichlet-vae-topic-models}} are directly used. The baselines of GSM and NVCTM are implemented by us based
on the code of NVDM, where the length of CTF in NVCTM is set to $10$ according to the preliminary experiments.
For our model, we use the widely adopted pre-trained word embedding from Glove \cite{Glove}, and the embedding size is $300$. All the models are trained alternatively by Adam optimizer with a learning rate of $1e^{-5}$ and a batch size of $64$. In the task of topic discovery, $ perplexity = \mathrm{exp} \{-\frac{1}{D}\sum_{d=1}^{D} \frac{1}{N_d} \sum_{i=1}^{N_d} \log p(w_{d,i})\}$ is used to evaluate the generalization performance of models on the testing set, where $D$ is the number of documents, $N_d$ is the number of words in document $d$ and $p(w_{d,i}) $ is the log-likelihood of model on word $w_i$ in document $d$. To evaluate the quality of discovered topics, we also use the normalized pointwise mutual information (NPMI) \cite{NPMI1} as the metric. The averaged values of NPMI on the top $5$, $10$, and $15$ words for all topics is computed as the final results. For the task of text classification, we use the topic vector of each document generated by convergent models as the input of a classifier. MLPClassifier from scikit-learn\footnote{\url{https://scikit-learn.org/stable/modules/classes.html}} is chosen as the classifier in this study and accuracy is used as the metric. For each task, the topic numbers are set to $25$, $50$, and $100$. We denote our models with Gaussian distribution and Gaussian mixture distribution in the decoder as CRNTM\_GD and CRNTM\_GMD, respectively. Unless explicitly specified, the number of Gaussian components is set to $25$ for CRNTM\_GMD. The source code, detailed parameter settings, and complementary results of our models can be found at Github\footnote{\url{https://github.com/Deloris-NLP/CRNTM}}.

\subsection{Comparison with Baselines}
Table \ref{tab:ppx} presents the test document perplexities of all models, from which we can observe that our CRNTM\_GD and CRNTM\_GMD achieve the best results in most cases. Specifically, TMN performs the best on Snippets. The reason may be that TMN is basically a supervised model for text classification and that the supervision from labels can help mining topics on a corpus consisting of less formal texts (i.e., Snippets). We also report the results of topic coherence in Table \ref{tab:coh}. It can be observed that both of our models are significantly better than all the baselines, which shows that they are able to discover more meaningful and interpretable topics. The performance comparisons for text classification are shown in Table \ref{tab:acc}. We can find that CRNTM\_GD and CRNTM\_GMD obtain competitive performances when compared with the benchmark methods, which validates the effectiveness of our models on generating representative vectors for short text classification.

\begin{table}
  \caption{Perplexity results of different models on both datasets, where the best scores are boldfaced.}
  \label{tab:ppx}
\scriptsize
\begin{center}
  \begin{tabular}{c|ccc|ccc}
    \toprule
	\multirow{2}*{Model} & \multicolumn{3}{c|}{20NewsGroups}	& \multicolumn{3}{c}{Snippets}\\	\cline{2-4}\cline{5-7}	
	& $25$	& $50$	& $100$	& $25$	& $50$	& $100$ \\
\midrule
NVDM	&	802 	&	855 	&	871 	&	5144	&	5180	&	5328	\\
NVLDA	&	1046 	&	1252 	&	1153 	&	5336	&	5496	&	5374	\\
ProdLDA	&	1106 	&	1073 	&	1035 	&	5312	&	5379	&	5348	\\
GSM		&	949  	&	 922 	&	 943	&	5237	&	5295	&	5434	\\
TMN	&	1159 	&	1136 	&	1128 	&	\textbf{3177}	&	\textbf{3197}	&	\textbf{3236}	\\
NVCTM	&	758 	&	738 	&	744 	&	5090	&	5121	&	5136	\\
DVAE	&	1095	&	1066	&	1075	&	5090	&	5121	&	5136	\\
CRNTM\_GD	&	698	&	706	&	680	&	4822	&	4872	&	4861	\\
CRNTM\_GMD	&	\textbf{574}	&	\textbf{586}	&	\textbf{590}	&	4608	&	4695	&	4602	\\
\bottomrule
  \end{tabular}
\end{center}
\end{table}

\begin{table}
  \caption{Topic coherence results of different models on both datasets, where the best scores are boldfaced.}
  \label{tab:coh}
\scriptsize
\begin{center}
  \begin{tabular}{c|ccc|ccc}
    \toprule
	\multirow{2}*{Model} & \multicolumn{3}{c|}{20NewsGroups}	& \multicolumn{3}{c}{Snippets}\\	\cline{2-4}\cline{5-7}	
& $25$	& $50$	& $100$	& $25$	& $50$	& $100$ \\
\midrule
NVDM	&	0.041	&	0.061	&	0.053	&	0.068	&	0.067	&	0.069	\\
NVLDA	&	0.065	&	0.062	&	0.061	&	0.042	&	0.045	&	0.041	\\
ProdLDA	&	0.064	&	0.062	&	0.065	&	0.046	&	0.051	&	0.045	\\
GSM		&	 0.080 &	0.076 	&	 0.065	&	0.068	&	0.061	&	0.065	\\
TMN	&	0.031	&	0.051	&	0.042	&	0.043	&	0.025	&	0.029	\\
NVCTM	&	0.022	&	0.017	&	0.014	&	0.052	&	0.051	&	0.055	\\
DVAE	&	0.065	&	0.075	&	0.069	&	0.039	&	0.052	&	0.040	\\
CRNTM\_GD	&	0.065	&	0.077	&	0.069	&	0.075	&	0.076	&	0.074	\\
CRNTM\_GMD	&	\textbf{0.088}	&	\textbf{0.081}	&	\textbf{0.079}	&	\textbf{0.082}	&	\textbf{0.084}	&	\textbf{0.085}	\\
\bottomrule
\end{tabular}
\end{center}
\end{table}

\begin{table}
  \caption{Classification accuracies of different models on both datasets, where the best scores are boldfaced.}
  \label{tab:acc}
\scriptsize
\begin{center}
  \begin{tabular}{c|ccc|ccc}
    \toprule
	\multirow{2}*{Model} & \multicolumn{3}{c|}{20NewsGroups}	& \multicolumn{3}{c}{Snippets}\\	\cline{2-4}\cline{5-7}	
& $25$	& $50$	& $100$	& $25$	& $50$	& $100$ \\
\midrule
NVDM	&	0.64	&	0.64	&	0.67	&	0.15	&	0.17	&	0.16	\\
NVLDA	&	0.40	&	0.45	&	0.42	&	0.12	&	0.13	&	0.13	\\
ProdLDA	&	0.43	&	0.44	&	0.40	&	0.14	&	0.14	&	0.15	\\
GSM		&	 0.45 	&	0.46 	&	0.45 	&	0.11	&	0.12	&	0.11	\\
TMN	&	0.40	&	0.48	&	0.51	&	0.15	&	0.16	&	0.13	\\
NVCTM	&	0.64	&	0.64	&	0.65	&	\textbf{0.16}	&	\textbf{0.18}	&	\textbf{0.18}	\\
DVAE	&	0.32	&	0.37	&	0.34	&	0.08	&	0.09	&	0.06	\\
CRNTM\_GD	&	0.64	&	\textbf{0.65}	&	\textbf{0.68}	&	0.15	&	0.16	&	0.14	\\
CRNTM\_GMD	&	\textbf{0.69}	&	\textbf{0.65}	&	0.66	&	\textbf{0.16}	&	0.16	&	0.17	\\
\bottomrule
\end{tabular}
\end{center}
\end{table}

\begin{figure*}[!htb]
	\centering
	\subfigure[Top 15 words in CRNTM\_GD, and $T1$, $T2$, $T3$, $T4$ are topic centroids.]{\label{fig:gt1}  \includegraphics[scale=0.20]{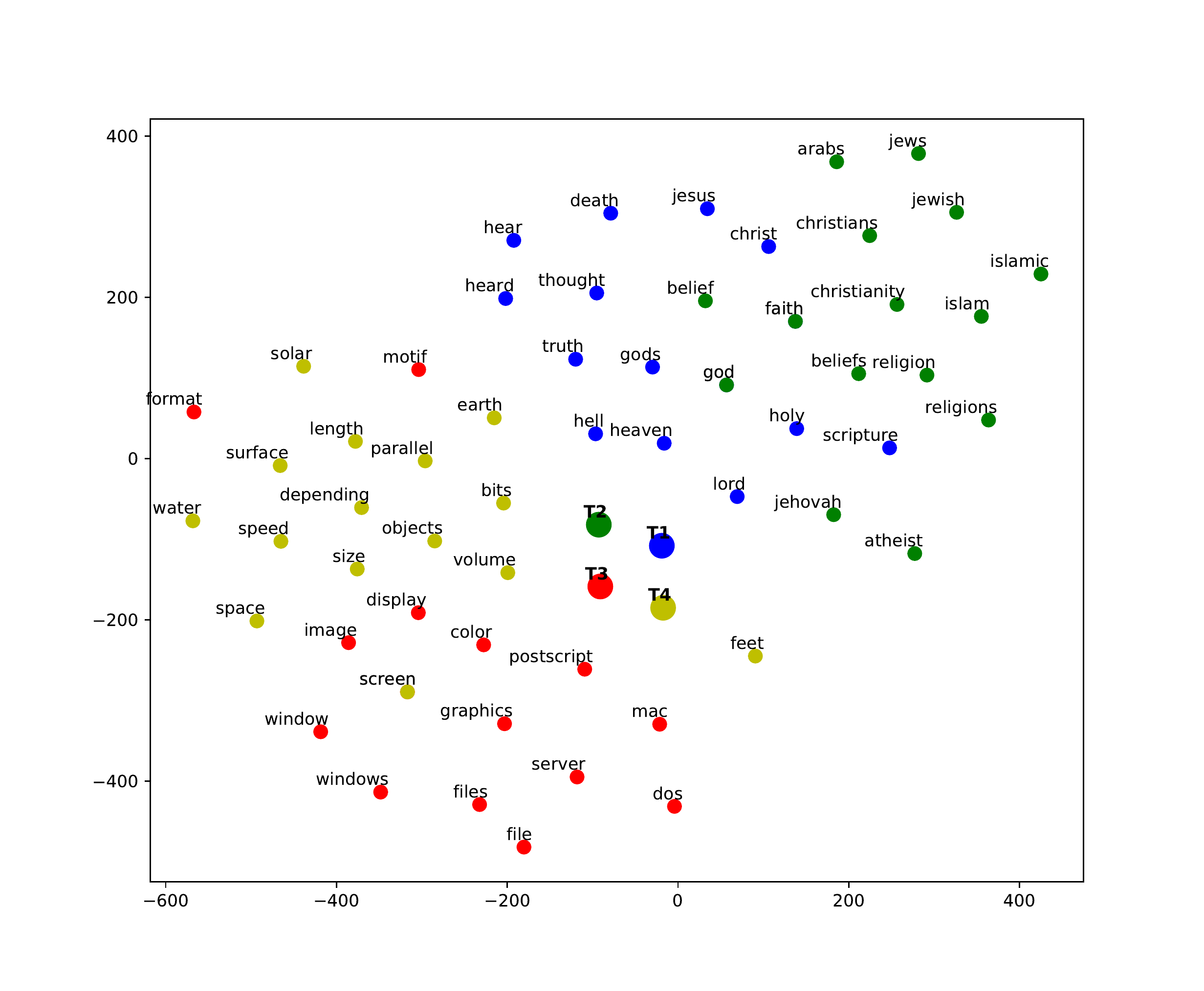}}
	\hspace{0.3cm}
	\subfigure[Top 15 words in CRNTM\_GMD. $Tk$:$m$ is the $m$th centroid of topic $k$. ]{\label{fig:gt2}  \includegraphics[scale=0.20]{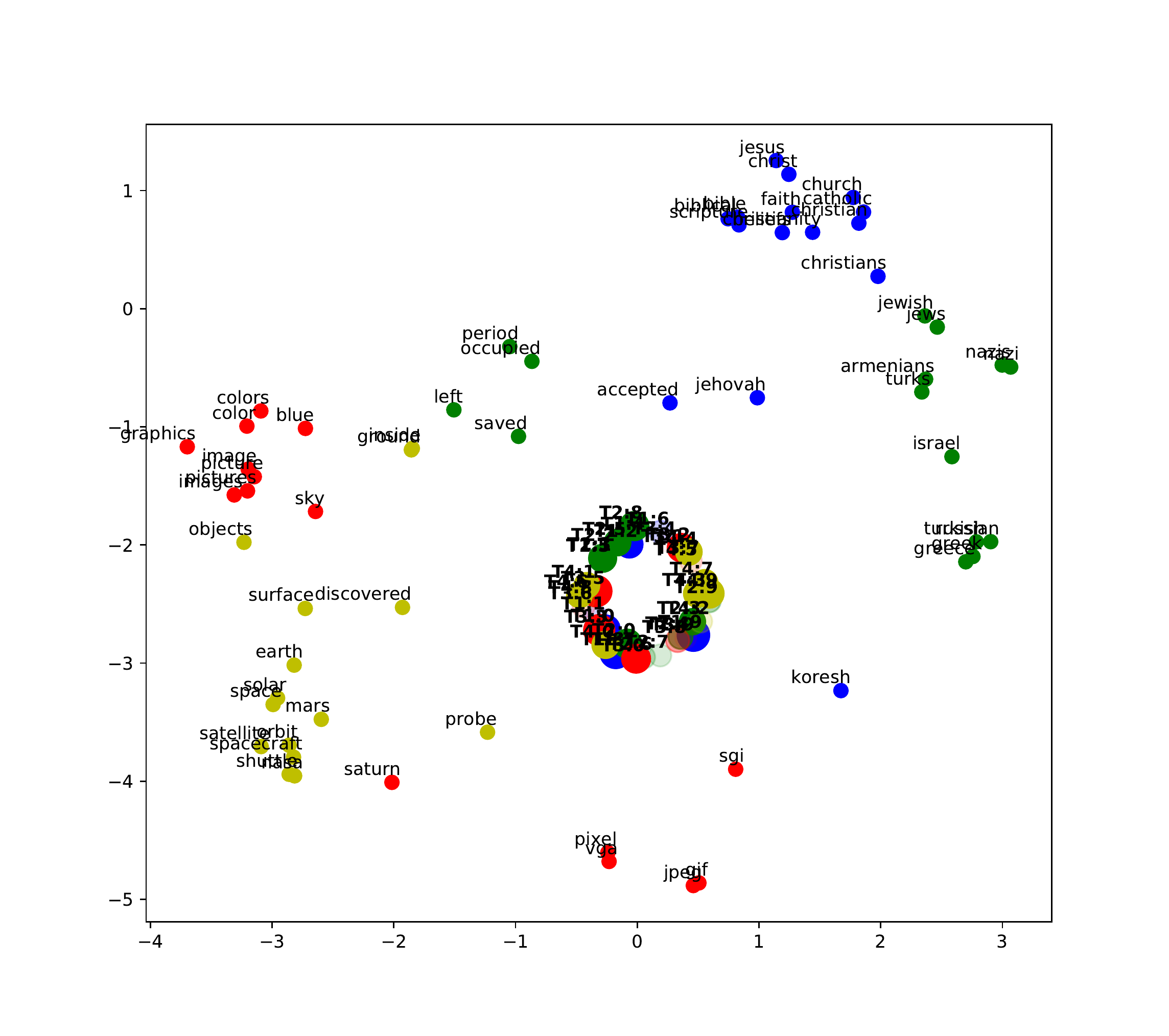}}
	\hspace{0.3cm}
	\subfigure[Topic centroids and their probabilities in CRNTM\_GMD. ]{\label{fig:gt3} \includegraphics[scale=0.20]{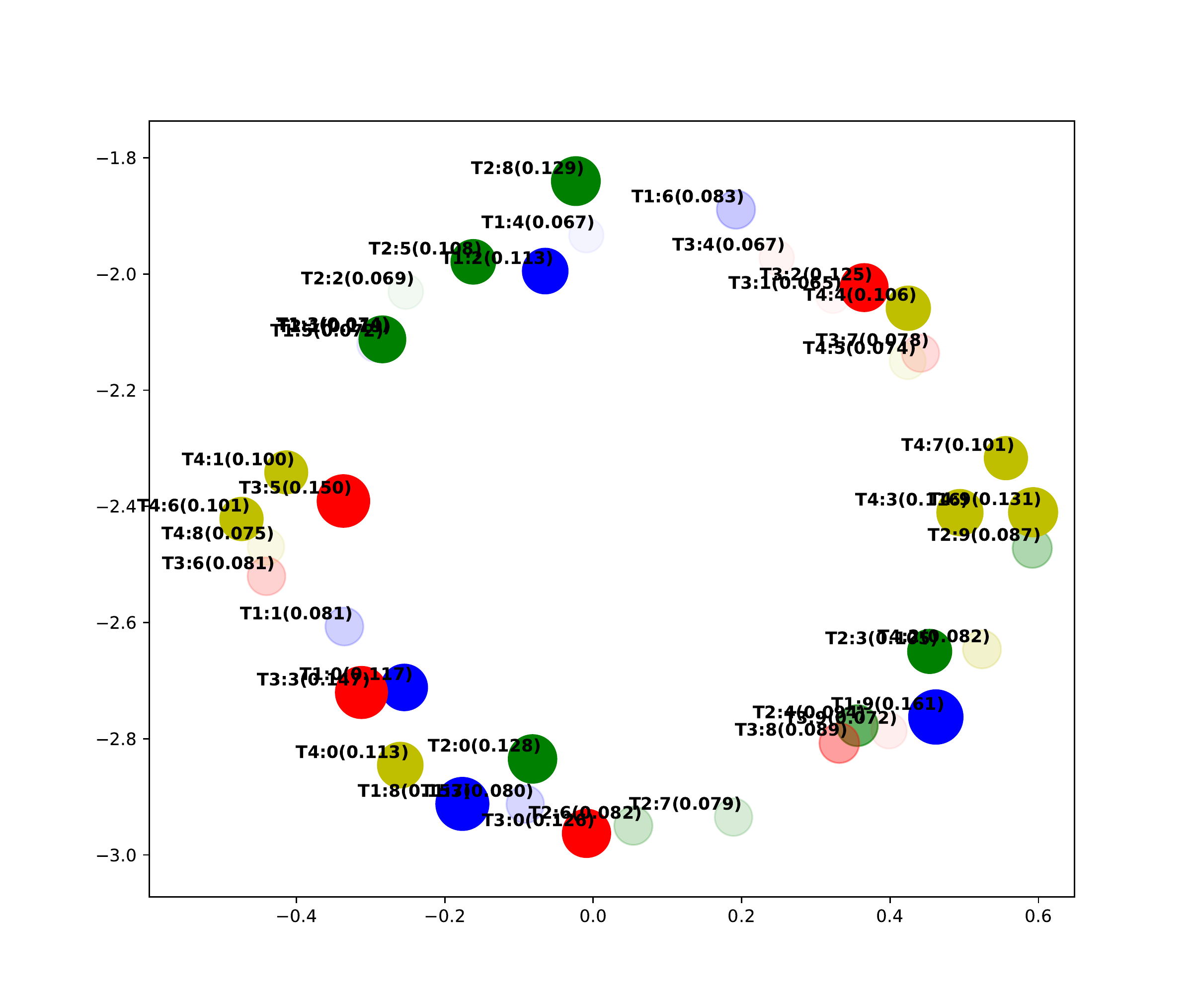}}
	\caption{Characteristics of 4 representative topics generated by our models on 20NewsGroups.}
\end{figure*}

\subsection{Evaluation on Gaussian Decoder via Topic Visualization}
To investigate the quality of topics discovered by our models, we report top 15 words of 4 representative topics and visualize these topics by their embedding vectors using 20NewsGroups. Particularly, we extract $\mu_k$ of Gaussian distributions as the topic centroid and utilize t-SNE \cite{maaten2009} for visualization. Topic visualization of the results in CRNTM\_GD is depicted in Figure \ref{fig:gt1}. The points with different colors indicate different topics, and the centroid of topic $k$ is denoted as $Tk$. For the convenience of comparison, we manually annotate each topic by referring to the ground truth category. Accordingly, $T1$, $T2$, $T3$, and $T4$ in CRNTM\_GD are annotated as ``soc.religion.christian'', ``talk.politics'', ``comp.sys.ibm.pc.hardware'', and ``comp.graphics'', respectively. We can see that all top words of the same topics are close to each other and to the corresponding topic centroids in the continuous vector space. This validates that our Gaussian decoder can effectively capture the context information via word embeddings in mining topics. We also present 4 topics generated by CRNTM\_GMD whose semantics are similar to those in CRNTM\_GD to verify the effectiveness of Gaussian mixture distributions. Topic visualization of the results in CRNTM\_GMD is shown in Figure \ref{fig:gt2}. For clarity, the coefficients of Gaussian components are indicated by different point sizes and shades of colour. The bigger the points are and the stronger the color is, the higher coefficient of the corresponding Gaussian components is. The topic centroids and their probabilities are detailed in Figure \ref{fig:gt3}. We can observe that for topic $T3$ named as ``comp.graphics'', the main components such as $T3$:$5$, $T3$:$3$, and $T3$:$0$ are close to the cluster of red points, while $T2$:$8$ and $T2$:$5$ are close to sub-clusters of top words in $T2$.

To make a comprehensive comparison, we present the results of all models on generating topic ``soc.religion.christian'' in Table \ref{tab:topwords}. It can be observed that our models can discover quite meaningful topics. We can also observe that the numbers of semantically irrelevant words of TMN and NVCTM are more than other models, which is consistent to the topic coherence results in Table \ref{tab:coh}.

\begin{table}[h]
  \caption{Top 10 words of manually labeled topic ``soc.religion.christian'' from all models on 20NewsGroups, where irrelevant words are underlined.}
  \label{tab:topwords}
\scriptsize
\begin{center}
  \begin{tabular}{c|l}
    \toprule[1.5pt]
Model	&	Top words \\
\midrule
	\multirow{2}*{NVDM}	&	god	sin	\underline{scsi}	bible	jesus \underline{rutgers}	\\	
&	homosexuality	christian	\underline{ide}	christians		\\
\midrule
\multirow{2}*{NVLDA}	&	god	scsi	sin	\underline{drive}	jesus	\\	
&	bible	christian	christians	homosexuality	love	\\
\midrule
\multirow{2}*{ProdLDA}	&	god	christians	jesus	bible	doctrine \underline{interpretation}	\\	
&	belief	homosexuality	christianity	eternal		\\
\midrule
\multirow{2}*{GSM}	&	god	jesus	bible	christ	church	\\	
&	\underline{people}	christian	believe	christians	sin	\\
\midrule
\multirow{2}*{TMN}	&	sin	\underline{myers}	eternal	\underline{president}	\underline{mary}	\\	
&	god	heaven	christ	\underline{doctor}	\underline{jobs}	\\
\midrule
\multirow{2}*{NVCTM}	&	church	catholic	christians	\underline{magnus}	scripture	\\	
&	\underline{duke}	\underline{andrew}	\underline{turkey}	\underline{sex}	christianity	\\
\midrule
\multirow{2}*{DVAE}	&	jesus	scripture	christ	bible	doctrine	\\	
&	sin	christians	god	canon	homosexuality	\\
\midrule
\multirow{2}*{CRNTM\_GD}	&	jesus	god	christ	heaven	death	\\	
&	holy	truth	gods	faith	lord	\\
\midrule
\multirow{2}*{CRNTM\_GMD}	&	god	christians	bible	christ	jesus	\\	
&	sin	religion	church	lord	doctrine	\\
\bottomrule[1.5pt]
\end{tabular}
\end{center}
\end{table}

\subsection{Impact of Gaussian Mixture Numbers}
We further study the impact of the number of Gaussian components. Table \ref{tab:gaussian} presents the results of CRNTM\_GMD on 20NewsGroups when varying component numbers under 25 topics. We can observe that CRNTM\_GMD with more Gaussian components generally performs better than that with less ones, which demonstrates that a more sophisticated mixture possesses a stronger capacity of learning high quality topics. The best topic coherence and classification accuracy are obtained when the component number is set to 25, and a larger value may not further boost the model performance.

\begin{table}
\small
	\caption{Performance of CRNTM\_GMD with different Gaussian mixture numbers on 20NewsGroups, where the best results are boldfaced.}
	\label{tab:gaussian}
	\begin{center}
		\begin{tabular}{c|ccc}
			\toprule[1.5pt]
			M	&	Perplexity	&	Coherence	&	Accuracy	\\
			\midrule[1.2pt]
			5	&	634	&	0.060	&	0.65	\\
			10	&	616	&	0.071	&	0.66	\\
			15	&	597	&	0.081	&	0.68	\\
			20	&	588	&	0.084	&	0.68	\\
			25	&	574	&	\textbf{0.088}	&	\textbf{0.69}	\\
			30	&	574	&	0.080	&	0.66	\\
			35	&	\textbf{571}	&	0.081	&	0.66 \\
			\bottomrule[1.5pt]
		\end{tabular}
	\end{center}
\end{table}

\section{Conclusion} \label{sec:conclusion}
In this paper, we propose a Context Reinforced Neural Topic Model (CRNTM) to address the feature sparsity problem in short texts. By introducing a \emph{topic controller} to the inference network, CRNTM infers the topic for each word in a narrow range. Besides, pre-trained word embeddings are incorporated with multivariate Gaussian distributions or Gaussian mixture distributions into our model to enrich the context information of short messages. To quantitatively validate the effectiveness of CRNTM, we conduct various experiments on two benchmark datasets in terms of perplexity, topic coherence, and text classification accuracy. The results indicate that the proposed model largely improves the performance of topic modeling by enriching the context information effectively.

\bibliographystyle{acl_natbib}
\bibliography{emnlp2020}

\end{document}